\documentclass[prl]{revtex4}
\usepackage{amsmath}
\usepackage{latexsym}
\usepackage{amsfonts}
\begin{document}
\title{Anyonic Excitations in Fast Rotating Bose Gases{\footnote{Presented
at the School on Quantum Phase Transitions and Non-Equilibrium Phenomena in
Cold Atomic Gases, ICTP Trieste, 11-22 july 2005 -- e-mail:
wallet@ipno.in2p3.fr}}}
\author{J.C. \surname{Wallet}}
%\maketitle
%\begin{center}
\affiliation{Groupe de Physique Th\'{e}orique, Institut de Physique Nucl\'{e}aire, 
F-91406 Orsay CEDEX, France}
%\end{center}
\begin{abstract} 
The role of anyonic excitations in fast rotating harmonically trapped Bose
gases in a fractional Quantum Hall state is examined. Standard Chern-Simons
anyons as well as "non standard" anyons obtained from a statistical
interaction having Maxwell-Chern-Simons dynamics and suitable non minimal
coupling to matter are considered. Their respective ability to stabilize
attractive Bose gases under fast rotation in the thermodynamical limit is
studied. Stability can be obtained for standard anyons while for non
standard anyons, stability requires that the range of the corresponding
statistical interaction does not exceed the typical wavelength of the
atoms.\par 
\vskip 0,5 true cm
\pacs: {PACS numbers: 03.75.Lm, 73.43.-f}
\end{abstract}

\maketitle

%\strut\thispagestyle{empty}
%\pagebreak
%\setcounter{page}{1}

%tableofcontents
%listoffigures
%\listoftables
\section{Introduction.}
The experimental realization of Bose-Einstein Condensation (BEC) of atomic
gases \cite{BEC} has given rise to a rich variety of phenomena and motivated
numerous investigations focussed on ultracold atomic Bose gases in rotating
harmonic traps. BEC confined to two dimensions have been created
\cite{BEC2D} and their response under rotation has been studied. Basically,
the rotation of a BEC produces vortices in the condensate
\cite{VORT1},\cite{VORT2}. When the rotation frequency increases, the BEC
state is destroyed and for sufficiently high frequency, a state
corresponding to Fractional Quantum Hall Effect (FQHE) \cite{FQE} is expected to
possibly occur \cite{F10}. In particular, when the rotation frequency is tuned to the
characteristic frequency of the harmonic confining potential in the radial
plane, FQHE states have been predicted to become possible ground states for the
system \cite{WILK}. This observation has been followed by studies focused on the
FQHE for bosons with short range interactions \cite{WORKS1}, \cite{WORKS2}. It is known 
that FQHE states involve anyonic excitations \cite{LERDA}.
Starting from the standard Chern-Simons (CS) realization for anyons
\cite{LERDA}, it has been shown in \cite{FISCHER} that in the
thermodynamical limit, a 2D harmonically trapped rotating Bose gas with
{\it{attractive}} interactions can be stabilized (in a FQHE state) thanks to
its anyonic (quasiparticle) excitations \cite{F20}. \par
\vskip 0,5 true cm
The above mentionned description of anyons is not unique. An alternative realization has been
investigated in \cite{NOUS} where the minimal coupling of the statistical
gauge potential to matter, which is now constrained to have
Maxwell-Chern-Simons (MCS) dynamics, is supplemented by a Pauli-type
couling, as recalled below. While the effective Landau-Ginzburg (LG)
theories stemming from each anyon realization both reproduce the (so far
observed) low energy features of the Quantum Hall Fluids (QHF), slight
differences do exist between CS and MCS anyons. These latter stemm from a
statistical interaction having finite range (whereas CS interaction has zero
range) and have an additional attractive mutual interaction (not present for
CS anyons). I point out that the existence of this attractive
mutual interaction may have an impact on the ability of (MCS) anyonic
excitations to stabilize an attractive Bose gas in a FQHE state. In
particular, stability requires that the range of the MCS
statistical interaction does not exceed the typical wavelength for the
atoms. The description of QHF within the MCSLG theory stemming from the MCS
description of anyons is also discussed.\par
\section{Chern-Simons and Maxwell-Chern-Simons anyons.}
Consider an interacting Bose gas at zero temperature 
in a rotating harmonic trap with strong confinement in the direction of the 
rotation axis so that the system is actually two dimensional. The
Hamiltonian in the rotating
frame \cite{DALF} and the corresponding action quoted here for further convenience
are (for conventions, see \cite{FOOT1})
$$H=\sum_{A=1}^N{{1}\over{2m}}({\bf{p}}_A-{\bf{A}}({\bf{x}}_A))^2
+\sum_{A=1}^N{{m}\over{2}}(\omega_T^2-\omega^2)|{\bf{x}}_A|^2
+\sum_{A<B}V({\bf{x}}_{AB})+... \eqno(1),$$
$$S_0=\int_t(\sum_{A=1}^N{{m}\over{2}}{\dot{\bf{x}}_A}^2-
\sum_{A=1}^N{{m}\over{2}}(\omega_T^2-\omega^2)|{\bf{x}}_A|^2-\sum_{A<B}V({\bf{x}}_{AB}))
+\int_{x}{\bf{A}}({\bf{x}}).{\bf{J}}({\bf{x}},t)+...\eqno(2),$$
where
${\bf{J}}({\bf{x}},t)$$=$$\sum_A$${\dot{\bf{x}}_A(t)}\delta({\bf{x}}-{\bf{x}}_A(t))$, the 
external gauge potential $A_i({\bf{x}}_A)$$=$$m\omega\epsilon_{ij}x^j_A$ yields the
Coriolis force, ${{\omega}\over{2\pi}}$(resp. ${{\omega_T}\over{2\pi}}$) is
the rotation (resp. trapping) frequency), $V({\bf{x}}_{AB})$ is the 
two-body potential felt by the bosons in the plane. It can be well
approximated by $V({\bf{x}}_{AB})$$=$$g_2\delta({\bf{x}}_{AB})$ since the
scattering between ultracold bosons is dominated by the s-wave \cite{DALF}.
In (1)-(2), the ellipses denote possible
multibody interaction terms. We now consider the "limit"
$\omega\simeq\omega_T$ for which the trapping and centrifugal potential
(nearly) balance each other so that $W$ can be
neglected.\par
\vskip 0,5 true cm
The standard CS description of anyons \cite{LERDA} obtained from (1)-(2) is achieved by
minimally coupling to $S_0$ a statistical CS gauge potential $a_\mu$:
$S_0$$\to$$S_1$$=$$S_0$$+$$\int_x{{\eta}\over{4}}\epsilon_{\mu\nu\rho}a^\mu
f^{\nu\rho}$$+$$a_\mu J^\mu$ where the CS parameter $\eta$ is dimensionless,
$J_\mu$$=$$(\rho;{\bf{J}})$ with
$\rho$$=$$\rho({\bf{x}},t)$$=$$\sum_A$$\delta({\bf{x}}$$-$${\bf{x}}_A(t))$ and ${\bf{J}}$ as given above
($f_\mu$$\equiv$${{1}\over{2}}\epsilon_{\mu\nu\rho}f^{\nu\rho}$, 
$f_{\mu\nu}$$=$$\partial_\mu a_\nu$$-$$\partial_\nu a_\mu$), $f_0$ (resp. $f_i$)
is the statistical magnetic (resp. electric) field). The equations of motion
for $a_\mu$ take the field-current identity form
$f_\mu$$=$$-$${{1}\over{\eta}}J_\mu$ (i), insuring that particle-statistical magnetic
flux composite anyonic objects are formed and permit one to determine
$a_\mu$ in term of the particle current \cite{LERDA}. Particle-flux coupling responsible
for the  formation of the above composite quasiparticles leads to the
occurence of Aharonov-Bohm type interactions. The anyonic
character of the wave functions for the quasiparticles is determined by
the Aharonov-Bohm phase \cite{LERDA} $\exp(i\int_C{\bf{a}}.d{\bf{x}})$$=$$\exp
{{i}\over{\eta}}$ ($C$
is some closed curve) that is induced when one quasiparticle moves
adiabatically around another one, equivalent to 
a {\it{double}} interchange of identical quasiparticles in the wave
function. Then, the statistics of the quasiparticles is controlled by the
value of $\eta$: they have Fermi(resp. Bose) statistics
whenever $\eta$$=$${{1}\over{(2k+1)2\pi}}$ (resp.${{1}\over{(2k)2\pi}}$),
$k$$\in$$\mathbb{Z}$. The statistics is anyonic otherwise. Notice that, when $\eta$$=$
${{1}\over{(2k+1)2\pi}}$, the initial Bose statistics of the atoms is
converted into a Fermi statistics for the quasiparticle which corresponds to
the so called statistical transmutation \cite{LERDA}. This, plugged into a second quantized formalism,
leads to the usual CSLG action \cite{ZHANG} underlying  the main part of the analyzis 
presented in \cite{FISCHER}. Since the statistical interaction is mediated by
a CS gauge potential, its range is zero. \par
\vskip 0,5 true cm
An alternative realization for anyons has been proposed and discussed in
\cite{NOUS}. It is obtained by coupling a MCS
statistical gauge potential $a_\mu$ to $S_0$ through minimal and non minimal
(Pauli-type) coupling term, the strength of this latter being fixed to a
unique specific value \cite{NOUS}, namely: $S_0$$\to$$S_2$
$=$$S_0$$+$$\int_x$$-$${{1}\over{4e^2}}f_{\mu\nu}f^{\mu\nu}$$+$${{\eta}\over{4}}\epsilon_{\mu\nu\rho}a^\mu
f^{\nu\rho}$$+$$a_\mu J^\mu$$-$${{1}\over{\eta e^2}}f_\mu J^\mu$ ($e^2$ has mass dimension $1$), where 
the coupling constant of the Pauli-term (last term in $S_2$) 
has been already fixed to the above mentionned specific value. Then, the equations of 
motion for $a_\mu$ can be written as
$-$${{1}\over{e^2}}\epsilon_{\alpha\mu\rho}\partial^\mu(f^\rho+{{1}\over{\eta}}J^\rho)
$$+$$\eta(f_\alpha+{{1}\over{\eta}}J_\alpha)$$=0$ which is solved by (i).This 
latter observation has been used as a starting point to construct an
effective theory reproducing the usual
anyonic behaviour \cite{NOUS}. The Aharonov-Bohm phase defining the
statistics of the resulting quasiparticles still verifies $\exp(i\int_C{\bf{a}}.d{\bf{x}})$$=$$\exp
{{i}\over{\eta}}$ so that $\eta$ again controls the statistics.
However, anyons obtained through this construction have an additional attractive
contact mutual interaction (not present in the CS case) \cite{NOUS}. Combining $S_2$
with the second quantization machinery, one obtains the corresponding MCSLG action
$$S=\int_x 
i\phi^\dag{\cal{D}}_0\phi-{{1}\over{2m}}|{\cal{D}}_i\phi|^2-g_2(\phi^\dag\phi)^2
-U(\phi)-{{1}\over{4e^2}}f_{\mu\nu}f^{\mu\nu}+{{\eta}\over{4}}\epsilon_{\mu\nu\rho}a^\mu
f^{\nu\rho} \eqno(3),$$
$${\cal{D}}_0=\partial_0-ia_0+{{if_0}\over{\eta e^2}}\ ;\
{\cal{D}}_i=\partial_i-i(a_i+A_i)+{{if_i}\over{\eta e^2}}
\eqno(4a;b).$$
In (3), $\phi$($=$$\phi({\bf{x}},t)$) is the order parameter, $\phi^\dag\phi$$=$$\rho$, $U(\phi)$
denotes the LG potential for multibody
interactions, $U(\phi)$$=$$g_3|\phi|^6$$+$$...$ ($g_3$$>$$0$ as in
\cite{FISCHER}). The non minimal coupling 
terms have been included in the extended covariant
derivative (4). The statistical MCS gauge potential has a finite mass
\cite{DESER} $M=\vert\eta\vert e^2$. Correspondingly, the statistical interaction 
has a finite range $\Lambda_{st}$$=$${{{1}\over M}}$. The
CSLG action \cite{ZHANG} is obtained from (3)-(4) by taking the limit
$e^2$$\to$$\infty$. While CSLG action  is believed to encode
the (so far observed) low energy physics of QHF, it
appears that MCSLG action also reaches this goal and could
therefore be used as an alternative description of QHF. This is
discussed more closely at the end of this letter. For the moment, we assume that 
the system is in a FQHE state ($\eta$$\ne$$0$) described
the MCSLG action (3)-(4).\par
\vskip 0,5 true cm
The equations of motion derived from (3) are
$$i{\cal{D}}_0\phi+{{1}\over{2m}}{\cal{D}}_i{\cal{D}}_i\phi-2g_2(\phi^\dag\phi)\phi=
{{\delta U(\phi)}\over{\delta\phi^\dag}};
-{{1}\over{e^2}}\epsilon_{\alpha\mu\rho}\partial^\mu(f^\rho+{{1}\over{\eta}}{\cal{J}}^\rho)
+(f_\alpha+{{1}\over{\eta}}{\cal{J}}_\alpha)=0 \eqno(5a;b)$$
where ${\cal{J}}_0$$=$$\rho$, ${\cal{J}}_i$$=$
${{i}\over{2m}}(\phi^\dag{\cal{D}}_i\phi-({\cal{D}}_i\phi)^\dag\phi)$
and ${\cal{D}}_\mu$ is defined in (4) while anyonic configurations, on which
we now focus, are obtained from (5b) when 
$f_\mu$$=$$-$${{1}\over{\eta}}{\cal{J}}_\mu$ \cite{FOOOT2}. From the field's conjugate momenta
$\Pi_{\phi^\dag}$$=$$\Pi_{a_0}=0$, $\Pi_{\phi}$$=$$i\phi^\dag$,
$\Pi_{a_i}$$=$${{1}\over{e^2}}f_{0i}$$-$${{\eta}\over{2}}\epsilon_{ij}a^j$$+$
${{1}\over{\eta e^2}}\epsilon_{ij}{\cal{J}}^j$, one obtains the Hamiltonian
$$H=\int_{\bf{x}}
{{1}\over{2e^2}}(f_0+{{1}\over{\eta}}\rho)^2
+{{1}\over{2e^2}}\Theta f_i^2+{{1}\over{2m}}\vert
D_i\phi\vert^2+(g_2-{{1}\over{2\eta^2e^2}})\rho^2+U(\phi) \eqno(6)$$
where $D_i$$=$$\partial_i$$-$$i(a_i+A_i)$ and we 
defined $\Theta$$=$$1$$-$${{\rho}\over{m\eta^2e^2}}$. The positivity of the model 
requires $\rho$$\le$$ m\eta^2e^2$ \cite{LATINSKY}. Restoring $\hbar$
and $c$, this translates 
into $\rho$$\le$$\rho_{lim}$, $\rho_{lim}$$=$
$\vert\eta\vert({{\Lambda}\over{\Lambda_{st}}}){{1}\over{\Lambda^2}}$
where $\Lambda$$=$${{\hbar}\over{mc}}$ is a typical (de Broglie) wavelength for
the atoms and $\Lambda_{st}$$=$${{\hbar}\over{|\eta|{\tilde{\mu}} c}}$
($e^2$$=$${\tilde{\mu}}c^2$). This condition should be fulfilled by current experimental values,
provided $\Lambda_{st}$$\lesssim$${\cal{O}}(\Lambda)$, a condition that we now
assume \cite{F2}. Namely, for $^7$Li, $^{23}$Na, $^{87}$Rb, one obtains respectively 
$\rho_{lim}$$\simeq$$\vert\eta\vert({{\Lambda}\over{\Lambda_{st}}})$
$10^{29}$, $10^{30}$, $10^{31}$ cm$^{-2}$, which, for possibly reachable
current Quantum Hall states, should be 
larger by several orders of magnitude than the experimental values reached by the matter
density so that ${{\rho}\over{m\eta^2e^2}}$$\ll$$1$.\par
\vskip 0,5 true cm
Compared to the CSLG Hamiltonian \cite{FISCHER}, $H$ involves additional
contributions from the Maxwell part of
(3) and the non minimal coupling terms in (4). The interaction energy
$g_2\rho^2$ receives contributions from ${{1}\over{2e^2}}f_0^2$ and the Pauli-type coupling
in (4a): these latter combine to yield the first term in (6) together with the
additional attractive (magnetic) contribution $-$${{\rho^2}\over{2\eta^2e^2}}$ to 
$g_2\rho^2$. This, depending on the relative magnitude of $\Lambda$ and
$\Lambda_{st}$, may somehow alter the conclusion obtained in \cite{FISCHER}
from a CSLG
description of anyons about the ability of statistical interaction to
stabilize an attractive Bose gas, as we now show.\par  
\section{Stabilizing an attractive 2D Bose gas.}
We set $\phi$$=$$\gamma e^{i\varphi}$$\equiv$${\sqrt{\rho}}e^{i\varphi}$, ${\hat{a}}_i$$=$
$\partial_i\varphi$$-$$(a_i$$+$$A_i)$, $\eta$$>$$0$ (here, one has
$\nu$$=$$2\pi\eta$ where $\nu$$>$$0$ is the filling factor; see below). For
$N$ atoms in the trap, one has $\int_{\bf{x}}\phi^\dag\phi$$=$$N$. Using 
$f_\mu$$=$$-$${{1}\over{\eta}}{\cal{J}}_\mu$ and assuming that fields vanish at infinity so that boundary terms
disappear, the static energy stemming from (6) can be conveniently written as
$$H=\omega N+\int_{\bf{x}}{{1}\over{2m}}(\partial_i\gamma
-\epsilon_{ij}{{{\hat{a}}^j\gamma}\over{{\sqrt{\Theta}}}})^2
+(g_2-g_0)\rho^2+\int_{\bf{x}}m\eta^2e^2\sum_{k=3}^\infty(\eta e^2C_{k-1}+2\omega
C_k)({{\rho}\over{m\eta^2e^2}})^k
+U(\gamma) \eqno(7).$$
In (7), $\omega$$>$$0$, the positive constants $C_k$'s are given by
$(1$$-$${{\rho}\over{m\eta^2e^2}})^{{{1}\over{2}}}$$=$$1$$-$
$\sum_{k=1}^\infty C_k({{\rho}\over{m\eta^2e^2}})^k$
($C_k$$=$${{\Gamma(k-{{1}\over{2}})}\over{2k!{\sqrt{\pi}}}}$ where $\Gamma$
is the Euler function) and we
have defined
$$g_0=-{{\hbar^2}\over{2m\eta}}(1-{{\Lambda_{st}}\over{\Lambda}}
(1-{{\hbar\omega}\over{2mc^2}}))\simeq 
-{{\hbar^2}\over{2m\eta}}(1-{{\Lambda_{st}}\over{\Lambda}})\eqno(8)$$
where $\hbar$ and $c$ have been reinstalled and the rightmost relation
stemms from $\hbar\omega$$\ll$$mc^2$ which holds for current experimental 
values for $\omega$ (taking ${{\omega}\over{2\pi}}$$\sim$${\cal{O}}(10)$-${\cal{O}}(10^3)$ Hz as a
benchmark). In the same way, the term $\sim$$\omega C_k$ in (7) can also been neglected
(since $\hbar\omega$$\ll$$\eta{\tilde{\mu}} c^2$ in view of
$\Lambda_{st}$$\lesssim$${\cal{O}}(\Lambda)$).\par
The quartic interaction terms in (7) can be eliminated provided $g_2$ is
chosen to be
$$g_2=g_0 \eqno(9).$$
Then, neglecting in (7) the small terms of order $\sim$${\cal{O}}(\gamma^6)$ (and
higher) as in \cite{FISCHER}, the ground state of (7) is obtained for those
configurations satisfying
$(\partial_i\gamma$$-$$\epsilon_{ij}{{{\hat{a}}^j\gamma}\over{\Theta^{{{1}\over{2}}}}})$$=$$0$,
$i$$=$$1,2$. This, combined with (5a), (8), (9), further expanding the various
contributions depending on $\Theta$ in powers of ${{\rho}\over{m\eta^2e^2}}$ and using the fact that
${{\rho}\over{m\eta^2e^2}}$$\ll$$1$ yields
$$\partial_i^2\gamma-{{(\partial_i\gamma)^2}\over{\gamma}}=-\gamma(
{{\gamma^2}\over{\eta}}(1+{{\omega}\over{\eta e^2}})+2m\omega) \eqno(10a),$$
where, in the RHS, ${{\omega}\over{\eta e^2}}$$\ll$$1$ still holds so that
(10a) can be accurately approximated by
$$\partial_i^2\gamma-{{(\partial_i\gamma)^2}\over{\gamma}}=-\gamma(
{{\gamma^2}\over{\eta}}+2m\omega) \eqno(10b).$$
When $\omega$ goes to zero, the
solutions of (10b) reduces smoothly \cite{EZAWA} to the solutions of the Liouville equation
$$\Delta\ln\rho=-{{2}\over{\eta}}\rho \eqno(11).$$
These latter can be parametrized as
$\rho(z)$$=$$4\eta|h^\prime(z)|^2(1+|h(z)|^2)^{-2}$ ($z$$=$$x_1$$+$$ix_2$)
for any holomorphic function $h(z)$. The particular choice
$h(z)$$=$$(z_0/z)^n$ ($z$$=$$re^{i\theta}$), $n\in \mathbb{N}$, gives rise
to radially symmetric vortex-type solutions $\rho(r)$$=$${{4\eta
n^2}\over{r^2}}(({{r_0}\over{r}})^n+({{r}\over{r_0}})^n)^{-2}$ where $r_0$
is some arbitrary length scale and $\phi$$=$${\sqrt{\rho}}e^{i n\theta}$.\par
\vskip 0,5 true cm
Let us discuss the above analyzis. From (7)-(9), one realizes that the
initial two-body coupling term $\sim$$g_2$ can be compensated by the
statistical interaction.  In view of (7), for fixed $\eta$, 
the particular value $g_0$ (8) represents the limiting value for the two-body
coupling constant below which the Bose gas cannot be stabilized by the
statistical interaction. When this latter is ruled by a CS dynamics
for which $\Lambda_{st}$$=$$0$, $g_0$ is negative while the ground
state of the system is exactly described (neglecting the small $|\phi|^6$
terms in $U(\phi)$) by (10b) giving rise to non singular finite energy 
matter distribution \cite{EZAWA}. In that situation, one would conclude as in
\cite{FISCHER} that an
attractive Bose gas may be stabilized against collapse in the
thermodynamical limit by anyonic excitations 
stemming from a CS (zero range) statistical interaction. Notice that
\cite{FISCHER} can be viewed as a nice reinterpretation in the context of
the trapped atomic Bose gases of the initial work \cite{PIJACK}. This
conclusion is somehow altered when the statistical interaction has a MCS
dynamics as described by (3) corresponding now to a (non zero) finite range.
Indeed, in view of (8), (up to small terms
$\sim$$\hbar\omega$), the sign of $g_0$ depends now on the relative
magnitude of $\Lambda$ and $\Lambda_{st}$. Compared to the CS case, the limiting value of the
two-body coupling constant $g_0$ receives a positive
contribution stemming from the (statistical) magnetic energy and the
magnetic coupling in (4a) in addition to the (negative) contribution of the
kinetic energy that can be converted into interaction energy. As a
consequence, negative values for $g_0$
can only be obtained when $\Lambda_{st}$$<$$\Lambda$ while a statistical
interaction whose range $\Lambda_{st}$ exceeds slightly $\Lambda$ gives rise
only to positive values for $g_0$ (non singular finite energy matter distribution
for the ground state is now well 
approximated by the solutions of (10b), again up to small $\gamma^6$ terms). This can be rephrased by stating
that anyonic excitations stemming from a MCS interaction with finite range
$\Lambda_{st}$ cannot protect an attractive Bose gas from collapse (in the thermodynamical
limit) unless $\Lambda_{st}$$\lesssim$$\Lambda$.\par
\section{MCS Landau-Ginzburg description of Quantum Hall Fluids.}
QHF are usually described
in the low energy regime by a CSLG theory \cite{ZHANG}. It
turns out that QHF could alternatively be described by a MCSLG theory such as 
the one defined in (3)-(4). To make more explicit the comparition
to the CS case \cite{ZHANG}, it is convenient to add a chemical potential term to (3), namely
$S^{\prime}$$=$$S$$+$$\int_x$$\mu\phi^\dag\phi$ ($\mu$$>$$0$) and set $U(\phi)$$=$$0$ 
in all. (5a) becomes
$i{\cal{D}}_0\phi$$+$${{1}\over{2m}}{\cal{D}}_i{\cal{D}}_i\phi$$-$$2g_2(\phi^\dag\phi)\phi$$=$
$\mu\phi$ while the relevant anyonic configurations are still obtained
when $f_\mu$$=$$-$${{1}\over{\eta}}{\cal{J}}_\mu$ solving (5b). Then, the
equations of motion admit the uniform (constant) density solution
minimizing the energy $H^\prime$$=$$H$$-$$\int_{\bf{x}}\mu\rho$, given by
$$\phi={\sqrt{n_0}};\ {\bf{a}}+{\bf{A}}=0;\ a_0=0;\ 
n_0={{\mu}\over{{2\hat{g_2}}}} \eqno(12a;b;c;d),$$ 
where ${\hat{g_2}}$$=$$g_2-{{1}\over{2\eta^2e^2}}$ which is similar to the
usual uniform ground state solution supported by the CSLG
action for QHF\cite{ZHANG}. Recall that (12b) implies that the
external magnetic field $B$$=$$\epsilon_{ij}$$\partial^iA^j$ is screened by the statistical magnetic field
(which signals (the analogue of) a Meissner type effect). This, further
combined with $f_0$$=$$-$${{1}\over{\eta}}\rho$, yields
$\eta$$=$${{n_0}\over{B}}$ so that $\eta$ is related to the filling
factor $\nu$ by $2\pi\eta$$=$$\nu$. In particular, when statistical
transmutation occurs as expected in Hall systems, $\nu$ satisfies
$\nu$$=$${{1}\over{2k+1}}$, $k\in \mathbb{Z}$. Note that (12c) implies that the (analog of) the
electric field seen by the particles is zero. Furthermore, introducing
polar coordinates $(r,\theta)$ and setting
$\phi$$=$${\sqrt{\rho}}e^{in\theta}$,
${\bf{a}}$$=$${\bf{e}}_\theta{{\alpha(r)}\over{r}}$
(${\bf{e}}_\theta$$=$$(\sin\theta,$$-$$\cos\theta)$), $a_0$$=$$a_0(r)$, where
$n$$\in$$\mathbb{Z}$ is the winding number, one finds after some algebra
that $H^\prime$ admits static finite energy vortex solutions satisfying the
boundary conditions  
$$\lim_{r\to\infty}\rho=n_0;\
\lim_{r\to\infty}{{\alpha(r)}\over{r^2}}=m\omega;\ \lim_{r\to
0}\rho=0;\ \lim_{r\to0}\alpha(r)=-n \eqno(13a;b;c;d)$$ 
so that, in view of (13a), far away from
the vortex center $r$$=$$0$ the order parameter $\phi$ behaves as
$\phi$$\sim$${\sqrt{n_0}}e^{in\theta}$. The vortex charge is
given by $Q$$=$$\int_{\bf{x}}$$(\rho$$-$$n_0)$ where the second term
represents the external (background) charge which must be substracted
\cite{EZAWA}. This, combined with
$n_0$$=$$\eta B$, $f_0$$=$$-$${{\rho}\over{\eta}}$ and the above boundaries
yields $Q$$=$$2\pi\eta n$. Therefore, when statistical transmutation occurs
$$Q={{n}\over{2k+1}} \eqno(14)$$
so that the  vortices can be interpreted as the fractionally charged Laughlin quasiparticles, as
it is the case in the CS description of QHF \cite{ZHANG}. Finally, expanding
$S^\prime$ around the (mean field) uniform solution (12) up to the quadatric
order in the fields fluctuations, fixing the gauge freedom and integrating
out the fluctuations of the statistical field gives rise to the low energy
effective action for the matter degree of freedom from which the dispersion
relation for the system is easily extracted. It can be written in obvious
notations as
$$\omega_0^2({\bf{p}})={{{\bf{p}}^4}\over{4m^2}}(1-2\Theta_0
[1-2({{m-\mu}\over{\eta
e^2}})^2])
+{\bf{p}}^2({{\mu}\over{m}}+\Theta_0+2\Theta_0^2({{m-\mu}\over{m}}))
+{{n_0^2}\over{m^2\eta^2}}(1-2\Theta_0^2) \eqno(15)$$
with $\Theta_0$$=$${{n_0}\over{m\eta^2e^2}}$ indicating that the system has
a gap (last term in (15)), as it could have been expected, so that the fluid
described by $S^\prime$ is incompressible. Notice that the terms involving
$\Theta_0$ can be safely neglected, since $\Theta_0$$\ll$$1$
holds in the present situation, so that (15) reduces to $\omega_0^2({\bf{p}})$$=$
${{{\bf{p}}^4}\over{4m^2}}$$+$${\bf{p}}^2({{\mu}\over{m}})$$+$
$+$${{n_0^2}\over{m^2\eta^2}}$.\par 
\section{Conclusion.}
The role of anyonic excitations in fast rotating harmonically trapped Bose
gases in a fractional Quantum Hall state has been examined. Standard CS
anyons as well as "non standard" anyons obtained from a statistical
interaction having MCS dynamics and suitable modified coupling to matter
have been considered. Their respective ability to stabilize attractive Bose
gases under fast rotation in the thermodynamical limit has been studied.
Stability can be obtained for standard anyons \cite{FISCHER} while for non
standard anyons, stability requires that the range of the corresponding
statistical interaction does not exceed the typical wavelength for the
atoms.\par
A part of the material involved in this note has been obtained in
collaboration with A. Lakhoua, M. Lassaut and T. Masson. I am grateful to
the organizers of the "School on Quantum Phase Transitions and Non-Equilibrium Phenomena in
Cold Atomic Gases" for invitation. Hospitality of ICTP Trieste is also
acknowledged.


\begin{thebibliography}{30}
\bibitem{F10}: When the trapping and centrifugal potential balance each
other, the bosons feel only the Coriolis force. Then, the system is formally
equivalent to that of 2D (bosonic) particles in a magnetic field,
corresponding to the condition for occurence of Quantum Hall regime.
\bibitem{FOOOT2}: It can be realized that the general (static) solutions of
(5b) such that $f_\mu$$\ne$$-$${{1}\over{\eta}}{\cal{J}}_\mu$ have infinite
energy.
\bibitem{F2}: ${{\Lambda}\over{\Lambda_{st}}}$$\ll$$1$ corresponds to a statistical
interaction with long range which seems to be unlikely. A large
$\Lambda_{st}$ would correspond either to
the small $e^2$ regime  where MCS theory enters a strong coupling regime 
for which a description of anyons as given by (3) is questionable or to a
fractional Quantum Hall state with $\eta$$\ll$$1$, e.g. state with extremely
large denominator.
\end{thebibliography}
\end{document}